\documentclass[5p,times]{elsarticle}

\usepackage{lineno,hyperref}

\usepackage{amssymb}
\usepackage{booktabs}
\usepackage[justification=centering]{caption}
\usepackage{colortbl}
\usepackage{csquotes}
\usepackage{enumitem}
\usepackage{float}
\usepackage{graphicx}
\usepackage[utf8]{inputenc}
\usepackage{multirow}
\usepackage{pgfgantt}
\usepackage{pgfplots}
\usepackage{pgfplotstable}
\usetikzlibrary{patterns}
\pgfplotsset{compat=1.9}
\usepackage{seqsplit}
\usepackage{subcaption}
\usepackage{threeparttable}
\usepackage{xcolor}

\journal{arXiv}

\begin{document}

\begin{frontmatter}

\title{From COBIT to ISO 42001: Evaluating Cybersecurity Frameworks for Opportunities, Risks, and Regulatory Compliance in Commercializing Large Language Models}

\author[1]{Timothy R. McIntosh\corref{mycorrespondingauthor}}
\cortext[mycorrespondingauthor]{Corresponding author}
\author[2]{Teo Susnjak}
\author[2]{Tong Liu}
\author[3]{Paul Watters}
\author[4]{Raza Nowrozy}
\author[5]{Malka N. Halgamuge}

\address[1]{Cyberoo Pty Ltd, Surrey Hills, NSW, Australia}
\address[2]{Massey University, Auckland, New Zealand}
\address[3]{Cyberstronomy Pty Ltd, Ballarat, VIC, Australia}
\address[4]{Victoria University, Melbourne, VIC, Australia}
\address[5]{RMIT University, Melbourne, VIC, Australia}

\begin{abstract}
	This study investigated the integration readiness of four predominant cybersecurity \textit{Governance, Risk and Compliance} (GRC) frameworks - NIST CSF 2.0, COBIT 2019, ISO 27001:2022, and the latest ISO 42001:2023 - for the opportunities, risks, and regulatory compliance when adopting \textit{Large Language Models} (LLMs), using qualitative content analysis and expert validation. Our analysis, with both LLMs and human experts in the loop, uncovered potential for LLM integration together with inadequacies in LLM risk oversight of those frameworks. Comparative gap analysis has highlighted that the new ISO 42001:2023, specifically designed for \textit{Artificial Intelligence} (AI) management systems, provided most comprehensive facilitation for LLM opportunities, whereas COBIT 2019 aligned most closely with the impending European Union AI Act. Nonetheless, our findings suggested that all evaluated frameworks would benefit from enhancements to more effectively and more comprehensively address the multifaceted risks associated with LLMs, indicating a critical and time-sensitive need for their continuous evolution. We propose integrating human-expert-in-the-loop validation processes as crucial for enhancing cybersecurity frameworks to support secure and compliant LLM integration, and discuss implications for the continuous evolution of cybersecurity GRC frameworks to support the secure integration of LLMs.
\end{abstract}

\begin{keyword}
cybersecurity frameworks \sep large language models \sep risk management \sep AI governance \sep NIST CSF \sep COBIT \sep ISO 27001 \sep ISO 42001

\end{keyword}

\end{frontmatter}

\section{Introduction}
\label{sec:Introduction}
Cybersecurity frameworks, such as the \textit{National Institute of Standards and Technology} (NIST) \textit{Cybersecurity Framework} (CSF), \textit{Control Objectives for Information and Related Technology} (COBIT), and \textit{International Organization for Standardization} (ISO) 27001 and 42001, are indispensable templates for diverse organizational sectors, with their dominance supported by recent industry reports and surveys \cite{kure2022integrated,sulistyowati2020comparative,tissir2021cybersecurity}. The NIST CSF 2.0 (\cite{nist2023csf20}), for instance, garners substantial endorsements both domestically and abroad, highlighted by its extensive academic citations \cite{dedeke2017cybersecurity,tissir2021cybersecurity}. Similarly, ISO 27001:2022 (\cite{iso27001}) boasts over 40,000 global certifications, emphasizing its stature in information security management \cite{hsu2016impact,mirtsch2020exploring}. COBIT 2019 (\cite{isaca2019cobit}), validated by several surveys, continues to be a premier choice for IT governance professionals \cite{atrinawati2021assessment,febriyani2022design,nugraheni2022adopting}. Recently introduced in December 2023, the ISO 42001:2023 (\cite{iso42001}) sets forth requirements for establishing and improving an \textit{Artificial Intelligence} (AI) Management System, yet there has been no systematic academic analysis of its advantages and disadvantages due to its novelty. Incorporating structured cyber risk navigation and enabling organizations to create custom cybersecurity governance tools, these frameworks continually evolve through a feedback-based approach of frequent revisions and ongoing collaborations with both public and private sectors, thereby addressing the dynamic technological landscape and emerging threats \cite{kure2022integrated,sulistyowati2020comparative,tissir2021cybersecurity,yusif2021conceptual}. Operationally guiding a range of functions that span asset categorization, control selection, training, and audits, these frameworks not only set the industry benchmark and serve as critical tools that consultants frequently leverage to evaluate the robustness of their clients' cybersecurity strategies, but also act as foundational pillars for academics, thereby significantly contributing to the enhancement of cybersecurity workforce expertise  \cite{bozkus2018cyber,kure2022integrated,yeoh2022systematic}. However, these frameworks face challenges in adapting their comprehensive controls to specific organizational environments, resources, and risk appetites. The integration of newer technologies, such as cloud computing and AI, adds to the complexity, necessitating advanced revisions to these frameworks \cite{kabanda2018exploring,radanliev2018integration,tawalbeh2020iot,tissir2021cybersecurity,tvaronavivciene2020cyber}. The cybersecurity landscape is slowly being transformed by the incorporation of \textit{Large Language Models} (LLMs), a change that Deloitte\footnote{https://www2.deloitte.com/uk/en/pages/deloitte-analytics/articles/embedding-controls-and-risk-mitigations-throughout-the-generative-ai-development-lifecycle.html} and KPMG\footnote{https://kpmg.com/xx/en/blogs/home/posts/2023/02/all-eyes-on-transforming-the-audit-with-ai.html} have already embraced, enhancing cybersecurity audit and operation capabilities and offering new opportunities for innovation in policy and compliance \cite{darraj2019artificial,mcintosh2023google,mcintosh2023harnessing,mcintosh2024inadequacies,yang2023dawn}. With over 92\% of Fortune 500 companies utilizing the OpenAI platform\footnote{https://www.cnbc.com/2023/11/06/openai-announces-more-powerful-gpt-4-turbo-and-cuts-prices.html}, LLMs are influencing corporate practices across various sectors, not just within cybersecurity. Nonetheless, LLMs can introduce their own set of challenges, particularly the risk of generating unreliable or `hallucinated' content, complicating their integration into existing cybersecurity measures \cite{ji2023survey,kaur2023artificial,mcintosh2023google,mcintosh2023harnessing}.

The ongoing discussion within the research community has been critical of the practicality and scientific underpinning of prevailing cybersecurity frameworks. Critics have pointed out the lack of empirical evidence to substantiate the effectiveness of these frameworks in enhancing security outcomes \cite{bayuk2013security,katina2021complex,malaivongs2022cyber,manuel2022cybertomp,paskauskas2022enisa}. They highlight the frameworks' propensity for detailed taxonomies of controls over actionable guidance for organizational-specific risk profiles \cite{argyridou2023cyber,bozkus2018cyber}. Further, it is argued that the frameworks do not sufficiently account for the multi-disciplinary nature required to address complex socio-technical challenges, often presenting a limited technical compliance viewpoint \cite{bayuk2013security,ekambaranathan2023can,shim2020internet}. Notably, existing frameworks have been found lacking in their coverage of new technologies, such as cloud computing and IoT \cite{darraj2019artificial,kaur2023artificial}. Conversely, supporters of these frameworks suggest they play a crucial role in raising awareness, unifying industry language, and embodying agreed-upon best practices, despite not ensuring security \cite{cho2015cyberphysical,hajny2021framework,manuel2022cybertomp}. Acknowledging these limitations, there is a concerted effort in recent research to augment cybersecurity frameworks by incorporating insights from the domains of security economics \cite{ekelund2019cybersecurity,radanliev2018integration,rathod2017novel}, behavioral psychology \cite{king2018characterizing,maalem2020review}, and system safety \cite{li2021comprehensive,taherdoost2022understanding}. The introduction of generative AI, such as LLMs, has fueled further debate, to include how to evolve these frameworks to safely utilize AI for security automation while managing associated risks, like model hallucinations \cite{mcintosh2023harnessing}. This synthesis of perspectives implies that while cybersecurity frameworks are beneficial starting points, they necessitate contextual adjustments and enhancements to drive substantive improvements in security programs.

This study assessed the readiness of four leading cybersecurity frameworks (\textit{i.e.}, COBIT 2019, ISO 27001:2022 (general purpose GRC), ISO 42001:2023 (AI Management System, or AIMS), and the draft NIST CSF 2.0) — chosen for their comprehensive coverage of \textit{Governance, Risk, and Compliance} (GRC) principles and their widespread adoption as one-stop shops for organizational GRC blueprints — in addressing the challenges and leveraging the opportunities presented by the integration of LLMs into cybersecurity operations. Amid a landscape where diverse bodies vie to set industry standards with frameworks that differ greatly in coverage, emphasis, and levels of abstraction, these were chosen for their robust encapsulation of GRC principles. Furthermore, the inherent abstractness and principle-based approach of these frameworks lend a degree of subjectivity to their interpretation, paralleling the diverse legal interpretations encountered in legislation. To address this, our study innovatively employs both LLMs and human experts in a loop, fostering a consensus-based interpretation to minimize disagreements. The study's motivation arose from NIST's call for feedback on its forthcoming NIST CSF 2.0, the recent release of ISO 42001:2023, and the draft of the \textit{European Union} (EU) AI Act, which is under consideration within the EU legislature\footnote{https://eur-lex.europa.eu/legal-content/EN/TXT/?uri=celex\%3A52021PC0206}. Our analysis focused on the secure, ethical use of generative AI, with an emphasis on LLMs, examining COBIT 2019, ISO 27001, ISO 42001, and the draft NIST CSF with equal rigor to deliver a comprehensive evaluation of their efficacy in the GRC context. Our research identified critical deficiencies in these frameworks, notably in the areas of human oversight, validation controls, and adherence to compliance—a crucial consideration in light of technologies like LLMs. Our comprehensive evaluation covered: (1) assessing the frameworks’ support for opportunities of adopting and integrating LLMs, (2) evaluating the inclusion of provisions for LLM risk mitigation, and human oversight and validation, and (3) determining the preparedness of the frameworks to align with the EU AI Act's main provisions, set to regulate the rapidly advancing generative AI industry that brought LLMs to global prominence in 2023. The intent of this research was not to directly instruct organizations on integrating LLMs into their GRC practices, but to stimulate informed discourse for timely enhancements to GRC frameworks. Such updates would bolster organizational reliance on these frameworks as they assimilate LLM technologies. Our findings have signaled an urgent need for framework modernization to address risks and compliance issues associated with emergent AI technologies, while capitalizing on the opportunities of their adoption and integration, through improved regulatory compliance and secure LLM guidelines. The analysis intends to ignite a vital, evidence-driven debate on the necessity for regular updates to cybersecurity standards, in the face of rapid technological evolution like LLMs.

This research makes several key contributions to the intersection of cybersecurity frameworks and LLM governance:
\begin{enumerate}[label=\arabic*)]
	\item We have provided one of the first academic evaluations of the preparedness of leading cybersecurity frameworks for integrating LLMs, revealing gaps in risk oversight.
	\item Our analysis of integration potential versus risk provisions has highlighted the need for a multi- dimensional approach as frameworks evolve to support LLMs.
	\item We have identified a lack of controls for managing LLM hallucination risks across frameworks, illuminating an issue that likely extends beyond the analyzed standards.
	\item Our findings have revealed the urgency of continuous evolution and timely version adoption for frameworks to address emerging technologies like LLMs, and in anticipation of regulatory shifts such as the forthcoming EU AI Act.
\end{enumerate}

The rest of this study is organized as follows: Section \ref{sec:RelatedWork} introduces related works. Section \ref{sec:methodology} explains our study methodology. Section \ref{sec:results} presents the results of our analysis. Section \ref{sec:discussion} discusses the key findings of our study. Section \ref{sec:future_directions} explores future research directions and proposes a roadmap to revise those cybersecurity frameworks. Section \ref{sec:conclusion} concludes this study.

\section{Related work}
\label{sec:RelatedWork}
This section examines studies related to the critical evaluation and enhancement of cybersecurity frameworks as well as the emerging threats posed by LLMs in the cybersecurity domain.

\subsection{Cybersecurity Framework Evaluation and Improvements}
There is a wealth of literature focused on the critical examination of mainstream cybersecurity frameworks, including the NIST CSF, COBIT, and ISO 27001. One recurrent theme centers on the limited empirical evidence supporting the real-world efficacy of these frameworks in improving security outcomes. For instance, a few studies (\textit{e.g.}, \cite{garvey2014analytical,gourisetti2020cybersecurity,gourisetti2021facility,hitchcox2020limitations,malaivongs2022cyber,renaud2021cyber}) found limited empirical data on cybersecurity framework effectiveness, and called for more rigorous, evidence-based studies on implementation impacts. Others have scrutinized the scientific validity of the risk management models embedded within frameworks. Some studies (\textit{e.g.}, \cite{argyridou2023cyber,bozkus2018cyber,hitchcox2020limitations,katina2021complex,kissoon2020optimum,manuel2022cybertomp,syafrizal2020analysis}) argued that a few main controls and recommendations of many cybersecurity frameworks lacked adequate theoretical and mathematical rigor or implementation practicality. Along similar lines, some studies (\textit{e.g.}, \cite{bayuk2013security,gourisetti2020cybersecurity,hitchcox2020limitations,syafrizal2020analysis}) highlighted scientific inconsistencies in some widely used frameworks, potentially suggesting that those frameworks were written based on the limited experiences of their authors or authoring groups.

Beyond scientific rigor, researchers have identified gaps in framework coverage and practical guidance, \textit{e.g.}, deficient or lacking security controls in the NIST CSF for emerging technologies \cite{darraj2019artificial,karie2021review,kaur2023artificial}, and limited actionable direction on control implementation \cite{goel2020prism}. To address these limitations, studies have recommended integrating complementary perspectives into frameworks, including security economics \cite{ekelund2019cybersecurity,radanliev2018integration,rathod2017novel}, behavioral psychology \cite{king2018characterizing,maalem2020review}, and system safety fields \cite{li2021comprehensive,taherdoost2022understanding}.

\subsection{LLM Threats in Cybersecurity}
The integration of LLMs into cybersecurity processes has prompted a surge in research focused on identifying and mitigating potential threats posed by these technologies. A critical issue is content hallucination, where LLMs generate plausible but factually incorrect information, which can have serious implications for cybersecurity, particularly in areas such as threat intelligence and incident response, where accuracy is paramount \cite{ji2023survey,mcintosh2023harnessing}. LLMs can create convincing yet entirely fabricated cyber threat reports, potentially leading to misinformed security measures \cite{gupta2023chatgpt,liu2023summary,weidinger2021ethical}. LLMs were found to have provided misleading information, which could lead to inadequate or erroneous vulnerability management, potentially leaving systems exposed to unaddressed security risks \cite{liu2023not,szabo2023new}. LLMs can be exploited to produce harmful or toxic outputs used for adversarial attacks, or be prompted to generate content that is seemingly benign but contains subtle manipulations intended to deceive or cause harm \cite{cheong2022envisioning,ji2023survey,qi2023visual}. LLMs can generate discriminatory biases within their outputs, which could inadvertently lead to biased cybersecurity practices, where such biases could manifest in security tools that rely on LLMs, potentially leading to unequal protection measures across different user groups \cite{burton2023algorithmic,zhang2023generative}.

To navigate these threats, the academic community has advocated for a human-centric approach to LLM governance in cybersecurity, which includes the development of frameworks that prioritize transparency, human oversight, and continuous evaluation of LLM outputs, to ensure that while LLMs can significantly contribute to cybersecurity efforts, they do so in a manner that is secure, ethical, and aligned with the overarching objectives of cyber defense strategies \cite{asad2023human,ukil2023knowledge}.

\section{Methodology}
\label{sec:methodology}
This section outlines our methodology, focusing on the criteria used to select cybersecurity frameworks and the analytical approach adopted for evaluation. Our methodology evaluated their effectiveness against both existing and emerging threats, particularly those introduced by LLMs. We began by examining the core elements of each framework, assessing their efficacy in the current threat landscape. Following this, we identified potential vulnerabilities to LLM threats, before making our final recommendations.

\subsection{Theoretical foundation}
\label{subsec:theoretical_foundation}
This research investigated the critical interplay between cybersecurity GRC and the evolving LLM landscape, emphasizing the essential role of understanding different challenges and prospects of LLMs in line with existing and expected regulations, thereby providing a holistic view of LLM integration into cybersecurity strategies.

\textbf{Governance:} Cybersecurity governance refers to the principles and practices designed to safeguard digital assets and data \cite{katina2021complex,yusif2021conceptual}. In the scope of this study, governance provides the blueprint to understand the structure and intent of cybersecurity frameworks. The main tenet here is that governance structures should be both robust and agile, especially in the face of novel AI-driven challenges such as LLMs \cite{asad2023human,ukil2023knowledge}.

\textbf{Risk Management:} The cornerstone of any robust cybersecurity strategy, risk management revolves around identifying, assessing, and addressing vulnerabilities and threats \cite{jarjoui2021framework,mcintosh2023harnessing}. LLMs introduce new avenues of risk: be it through generating malicious content or identifying gaps in security frameworks. Understanding this dynamic ensures a comprehensive evaluation of the frameworks under study.

\textbf{Compliance and Legislative Aspects:} The draft of the EU AI Act, anticipated to be ready by 2024, underscores the critical importance of compliance in the era of advanced AI \cite{dhirani2023ethical,khan2023embracing}. We believe the EU AI Act is likely to become the blueprint for other jurisdictions to propose their own AI regulations, akin to how the EU \textit{General Data Protection Regulation} (GDPR) has set the blueprint for others to enhance their Privacy Act. AI solutions, including LLMs, while not yet required to adhere strictly to the draft, would benefit from taking its main provisions into consideration early in their development roadmap, to prevent potential last-minute rushes or costly system revisions later on. 

\textbf{AI Ethics and Oversight:} As LLMs find utility in the research methodology and the EU AI Act draft gains prominence, ethical facets associated with LLM deployment warrant serious deliberation \cite{dhirani2023ethical,floridi2021ethical,weidinger2021ethical}. It extends beyond mere deployment, and upholds the principles of transparency in LLM processes, ensuring that LLMs remain accountable for their decisions, and upholding fairness, particularly when those models intersect with or influence cybersecurity protocols \cite{dhirani2023ethical,weidinger2021ethical}. Recognizing and addressing these ethical dimensions can solidify the credibility and trustworthiness of LLMs in cybersecurity contexts.

\subsection{Study design and validation}
\label{subsec:study_design}
\begin{figure*}[t!]
	\centering
	\includegraphics[width=\textwidth]{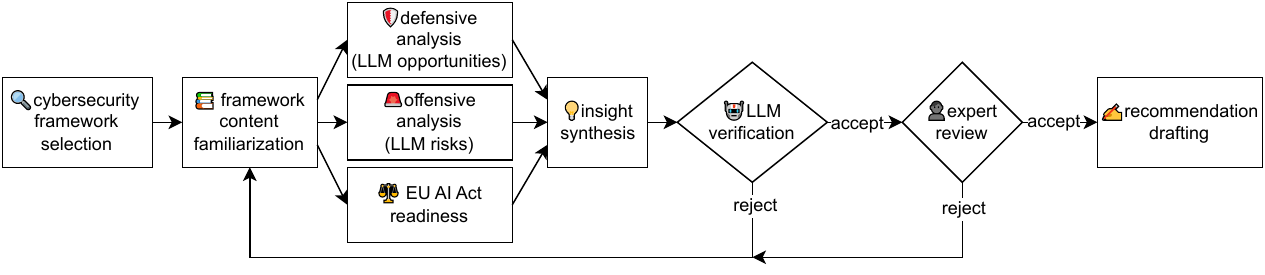}
	\caption{Flowchart of the analysis process with LLM and GRC experts in the loop}
	\label{fig:trifocal_flowchart}
\end{figure*}

This study adopted a thorough approach to assess cybersecurity frameworks, focusing on their engagement with LLMs within the advanced AI context, aiming to evaluate these frameworks' comprehensive preparedness for LLM-related threats and the forthcoming EU AI Act. The progression of our study is illustrated in Figure \ref{fig:trifocal_flowchart}. To ensure the maximum transparency, credibility, and reproducibility of our study, we exclusively used publicly available documents, specifically cybersecurity frameworks and the EU AI Act draft, without incorporating any customized data, to enable further scrutiny by other researchers for transparency and trustworthiness.

The steps of our study is as follows:

\begin{enumerate}[label=\arabic*)]
	\item \textit{Cybersecurity Framework Selection:} Selecting the appropriate cybersecurity GRC frameworks for evaluation. 
	
	\item \textit{Framework Content Familiarization:} An extensive analysis of the cybersecurity framework content. The lead author, proficient in cybersecurity GRC, ensures that this foundational analysis is robust for subsequent steps.
	
	\item \textit{Tri-Focal Analysis:}
	\begin{itemize}
		\item \textit{Defensive Analysis:} Evaluating the resilience of frameworks against LLM threats and their adaptability to harness LLM benefits, including:
		\begin{itemize}
			\item Provisions of the framework against LLM-generated malicious content.
			\item Mechanisms to harness the strengths of LLMs for enhanced cybersecurity.
		\end{itemize}
		
		\item \textit{Offensive Analysis:} Unearthing potential vulnerabilities in the frameworks due to LLM interactions, including:
		\begin{itemize}
			\item The ability of LLMs to produce misleading content.
			\item Identifying gaps within the framework that LLMs might exploit.
		\end{itemize}
		
		\item \textit{EU AI Act Readiness:} Evaluation of the cybersecurity framework in relation to the draft EU AI Act provisions.
	\end{itemize}
	
	\item \textit{Insight Synthesis:} Consolidation of the findings from the tri-focal analysis and synthesis of initial insights.
	
	\item \textit{LLM Verification Method:}
	Construction of synthesized insights, combined with relevant sections from the selected cybersecurity framework and supporting evidence, as prompts for OpenAI's ChatGPT-4 and Anthropic's Claude, to independently:
	\begin{itemize}
		\item Evaluate the validity of our initial analysis.
		\item Highlight any potential gaps or oversights.
	\end{itemize}
	Upon LLM feedback, if either ChatGPT-4 or Claude rejects our conclusions, or their reasoning does not align with our assessment, we restart our analysis. If they reject our analysis based on apparent hallucinations, we still restart, enhancing our evidence for better clarity. If both LLMs agree with our insights, we proceed to the expert review phase.

	\item \textit{Expert Review:}
	The LLM-validated synthesis undergoes scrutiny by two independent subject matter experts in cybersecurity GRC, ensuring they had no part in the initial insights synthesis to reduce bias.
	\begin{itemize}
		\item If the panel rejects, the lead author revisits content familiarization.
		\item If accepted, progression to recommendation drafting ensues.
	\end{itemize}

	\item \textit{Recommendation Drafting:} Drafting actionable recommendations to strengthen the cybersecurity frameworks based on insights after the validation process.
\end{enumerate}

\subsection{Framework selection}
\label{subsec:framework_selection}
All three cybersecurity frameworks considered in this research were sourced directly from their respective official websites to ensure authenticity, transparency, and accuracy of information. The selection was driven by the comprehensive coverage each framework provided across the three cardinal pillars of GRC: \textit{governance}, \textit{risk management}, and \textit{compliance}. Drawing our assessment to a close, it became evident from Table~\ref{tab:framework_selection} that the NIST CSF 2.0, COBIT 2019, and ISO 27001:2022 emerged as the most suitable choices for this study.

\begin{table}[t!]
	\centering
		\caption{Evaluation of Cybersecurity Frameworks\\ (\checkmark: comprehensive; $\bigtriangleup$: partial; *: selected)}
	\label{tab:framework_selection}
	\resizebox{\columnwidth}{!}{%
		\begin{tabular}{lcccc}
			\hline
			\multicolumn{1}{c}{\multirow{2}{*}{\textbf{Frameworks}}} & \multicolumn{3}{c}{\textbf{Cybersecurity GRC Blueprint Suitability}} & \multicolumn{1}{c}{\multirow{2}{*}{\textbf{\begin{tabular}[c]{@{}c@{}}Final \\ Selection\end{tabular}}} } \\
			\cline{2-4}
			& \textbf{Governance} & \multirow{2}{*}{\textbf{\begin{tabular}[c]{@{}c@{}} \\ Selection\end{tabular}}}  & \textbf{Compliance} & \\
			\hline
			NIST CSF 2.0 & \checkmark & \checkmark & \checkmark & * \\
			\arrayrulecolor{gray!40} \hline
			COBIT 2019 & \checkmark & \checkmark & \checkmark & * \\
			\hline
			ISO 27001:2022 & \checkmark & \checkmark & \checkmark & * \\
			\hline
			ISO 42001:2023 & \checkmark & \checkmark & \checkmark & * \\
			\hline
			CIS Controls &  & \checkmark &  &  \\
			\hline
			Australian Essential 8 &  & \checkmark &  &  \\
			\hline
			ITIL & $\bigtriangleup$ &  &  &  \\
			\hline
			GDPR &  &  & \checkmark &  \\
			\hline
			HIPAA &  &  & \checkmark &  \\
			\hline
			PCI DSS &  & $\bigtriangleup$ & \checkmark & \\
			\hline
			FAIR & $\bigtriangleup$ & \checkmark &  & \\
			\hline
			CIS RAM & $\bigtriangleup$ & \checkmark &  & \\
			\hline
			
			SABSA & \checkmark & \checkmark &  & \\ 
			\hline
			
			TOGAF & \checkmark & $\bigtriangleup$ & $\bigtriangleup$ & \\
			\hline
			MITRE ATT\&CK &  & \checkmark &  & \\
			\hline
			CSA Standards & $\bigtriangleup$ & $\bigtriangleup$ & $\bigtriangleup$ & \\ \arrayrulecolor{black}
			\hline		
		\end{tabular}
	}
\end{table}

To provide clarity on our selection rationale:
\begin{itemize}
	\item \textit{NIST CSF 2.0, COBIT 2019, ISO 27001:2022, and ISO 42001:2023}: These frameworks were selected due to their comprehensive coverage of all three GRC components of commercial LLMs, for their adaptability to evolving technological contexts, and for their widespread global acceptance.
	
	\item \textit{CIS Controls}: Though it offers vital technical guidelines, it does not possess the broad global adoption seen with our chosen frameworks.
	
	\item \textit{Australian Essential Eight}: This framework is predominantly tailored to technical standards on Microsoft platforms within Australia, limiting its global applicability.
	
	\item \textit{ITIL (Information Technology Infrastructure Library)}: A recognized global standard; however, its primary focus is IT service management, without fully encapsulating the broader spectrum of GRC.
	
	\item \textit{PCI DSS (Payment Card Industry Data Security Standard)}: Its primary emphasis is the security assurance of the payment card industry, making its scope more narrow compared to our selected frameworks.
	
	\item \textit{FAIR (Factor Analysis of Information Risk), and CIS RAM (Center for Internet Security Risk Assessment Method)}: While both tools emphasize risk assessment, they fall short in providing robust governance structures or clear compliance guidelines, making them less versatile in a GRC-focused approach.

	\item \textit{SABSA (Sherwood Applied Business Security Architecture)}: While recognized globally, its central thrust is on security architecture, diverging from the comprehensive GRC approach we sought.
	
	\item \textit{TOGAF (The Open Group Architecture Framework)}: An enterprise architecture methodology and framework, TOGAF ensures alignment between business and IT, providing strategic governance, efficient risk management, and broad compliance aspects.
	
	\item \textit{MITRE ATT\&CK (MITRE Corporation Adversarial Tactics, Techniques, and Common Knowledge)}: Primarily known as a knowledge base of attacker behaviors, it is increasingly referenced for cybersecurity risk management and governance. However, its primary focus isn't broad GRC, but it provides valuable insights into potential threats and techniques.
	
	\item \textit{CSA (Cloud Security Alliance) Standards}: While CSA provides best practices across governance, risk management, and compliance, its focus is primarily on cloud-centric environments, which limits its broader applicability in diverse technological contexts.

\end{itemize}

\subsection{Procedure for LLM-related analysis}
\label{subsec:analysis_procedure}
To ensure a systematic and rigorous assessment of cybersecurity frameworks concerning LLM integration, our evaluation divided the analysis into defensive and offensive categories:
\begin{enumerate}[label=\arabic*)]
	\item \textit{Defensive Analysis:}
	\begin{itemize}
		\item \textit{Identification of LLM Integration Potential}: We used the mapping rubric, we assessed each framework for processes amenable to LLM augmentation, correlating opportunities with known LLM capabilities to ensure feasible and beneficial integration.
		\item \textit{Assessment of Controls for LLM Risks}: We listed LLM-specific controls using the rubric as a guideline. Then, we assessed the ability of each control to handle LLM risks.
	\end{itemize}
	
	\item \textit{Offensive Analysis:}
	\begin{itemize}
		\item \textit{Framework Vulnerabilities to LLM Attacks}: We examined potential framework weaknesses using the rubric, focusing on areas LLMs could exploit. We also identified scenarios where LLMs may bypass regulations through generated content.
		\item \textit{Holistic Framework Gap Analysis}: We comparatively evaluated overall LLM readiness using the rubric, spotlighting areas needing more LLM-specific provisions.
	\end{itemize}
\end{enumerate}

\begin{table*}[t!]
	\centering
	\caption{Mapping Rubric for Framework Attributes, LLM Characteristics, and EU AI Act Readiness}
	\label{table:mapping_rubric}
	\begin{threeparttable} 
		\resizebox{\textwidth}{!}{%
			\begin{tabular}{|p{4cm}|p{4cm}|p{4cm}|p{7cm}|}
				\hline
				\textbf{Cybersecurity Framework Provisions} & \textbf{LLM Capabilities} & \textbf{LLM-related Risks}\tnote{a} & \textbf{EU AI Act Readiness} \\
				\hline
				Process automation \cite{alromaih2022continuous,maglaras2021digital} & Natural language understanding and generation \cite{min2023recent,yang2023dawn,zhang2022survey} & Misleading content generation \cite{ji2023survey,min2023recent} & Compliance (\textit{Article 13, 52}), transparency (\textit{Article 52-5}3), human oversight (\textit{Article 14, 29, 61, 63}). \\
				\hline
				
				Real-time analysis \cite{abie2019cognitive,akande2023cybersecurity,mcintosh2023harnessing} & Real-time data processing \cite{min2023recent,zhang2022survey} & Bias, unpredictability, latency in responses \cite{burton2023algorithmic,mcintosh2023harnessing,zhang2023generative} & Timely, unbiased AI system responsiveness (\textit{Article 9, 13, 14}). Real-time safeguards against risks (\textit{Article 5, 9, 62, 65}). \\
				\hline
				
				Data security and protection \cite{markopoulou2019new,sule2021cybersecurity} & Data analytics \cite{min2023recent,yang2023dawn,zhang2022survey}, image recognition and generation \cite{cheong2022envisioning} & Potential for data leakage \cite{montagna2023data,winograd2023loose}, visual deception, forensic unreliability \cite{cheong2022envisioning} & Robust protection against data breaches (\textit{Article 15, 70}). Ensuring AI data quality and integrity (\textit{Article 10. Annex IV-2, VII-4.3}). \\
				\hline
				
				Continuous monitoring and auditing \cite{gourisetti2021facility,malatji2019socio} & Continuous learning \cite{min2023recent,zhang2022survey} & Over-reliance on outdated training data \cite{singhal2023large} & Periodic AI assessment (\textit{Article 61, 84}). Addressing outdated data (\textit{Article 10, 43, 61}). \\
				\hline
				
				Incident response \cite{dykstra2022action,mcintosh2023harnessing} & Automated incident detection and reporting \cite{gupta2023chatgpt,iturbe2023artificial} & Misclassification of incidents \cite{rjoub2023survey} & Rapid AI-driven incident recognition (\textit{Article 62, 68}). Mitigate misclassification risks (\textit{Article 10, 13, 14, 15, 43, 61}). \\
				\hline
				
				Security awareness training \cite{khader2021cybersecurity,triplett2022addressing} & Adaptive training modules \cite{kasneci2023chatgpt} & Distorted reality focus \cite{ji2023survey,min2023recent}. Data profiling risks \cite{weidinger2021ethical} & AI transparency (\textit{Article 13}). Authenticity and accuracy (\textit{Article 15}). Profiling restrictions (\textit{Article 5, 52}). \\
				\hline
				
				Policy and compliance checks \cite{li2019investigating,mcintosh2023harnessing} & Automated policy drafting and checks \cite{mcintosh2023harnessing} & Introduction of non-compliant rules \cite{wang2023self} & Automated information verification (\textit{Article 60, 61, 64}). AI alignment with compliance (\textit{Article 8, 9, 16, 24-27}). \\
				\hline
			\end{tabular}
		}
		\begin{tablenotes} 
			\item[a] {\footnotesize including LLM-induced cybersecurity risks and inherent LLM risks}
		\end{tablenotes}
	\end{threeparttable} 
\end{table*}

The mapping rubric, central to our methodology, has been developed to effectively pair specific framework attributes with established LLM opportunities and potential risks, as presented in Table \ref{table:mapping_rubric}. Due to the abstract principle-based nature of those frameworks, to appraise their LLM-readiness, we employed a qualitative binary \enquote{pass/fail} criterion, where \enquote{pass} indicates the framework is LLM-ready, and \enquote{fail} suggests the opposite. The use of a binary \enquote{pass/fail} criterion is justifiable on several grounds:

\begin{itemize}
	\item \textit{Nature of Qualitative Research}: Qualitative research is inherently interpretative, focusing on understanding the complexity and context of a subject rather than reducing it to numbers \cite{fujs2019power}. We believe a binary \enquote{pass/fail} system aligns with this interpretative nature, providing a clear, dichotomous outcome that reflects a framework's readiness without the false precision of a numerical score.
	
	\item \textit{Complexity and Abstraction}: Cybersecurity frameworks are often characterized by their complexity and high level of abstraction, with provisions that cannot be easily quantified \cite{malatji2019socio}. We believe a granular scoring system could oversimplify these provisions, potentially misrepresenting the complex assessment required for LLM integration. A binary system, by contrast, acknowledges this complexity and avoids the pitfalls of over-simplification.
	
	\item \textit{Precedence in Qualitative Research}: Binary rating scales are not uncommon in qualitative evaluations, particularly in fields dealing with abstract concepts such as policy analysis and compliance assessments \cite{armenia2021dynamic,pipyros2018new}. For instance, in the evaluation of regulatory frameworks, binary outcomes are often preferred to indicate adherence or non-adherence to standards, as they facilitate clear decision-making and action \cite{armenia2021dynamic,malatji2019socio,pipyros2018new}.
\end{itemize}

Our approach, leveraging both human expertise and AI validation, provided a robust mechanism with a higher level of scrutiny and cross-validation than through human efforts alone, for ensuring the accuracy and integrity of our qualitative analysis. Our iterative process of human-LLM consensus served to balance the depth of human judgment with the breadth of the NLP analysis by LLM, as we integrated OpenAI's ChatGPT-4 and Anthropic's Claude AI into our analysis pipeline as follows:

\begin{itemize}
	\item \textit{Automated Validation Process}: Initially, the alignment of each framework with LLM capabilities and the EU AI Act provisions was independently assessed by the researchers. Then, to validate the analysis, the frameworks were processed through market-leading LLMs: ChatGPT-4, chosen for its advanced natural language prowess \cite{mcintosh2023harnessing}, and Claude, selected for its market-recognized reliability and robustness to hallucinations \cite{toufiq2023harnessing}. Both LLMs scrutinized the text of the frameworks and our initial assessments to pinpoint any discrepancies or aspects that may have been missed.
	
	\item \textit{Consistency Checks}: The AI systems were tasked with verifying the consistency of applying our mapping rubric. They cross-referenced the identified LLM capabilities and risks with the provisions of the frameworks to ensure a thorough and unbiased application of the rubric criteria.
	
	\item \textit{Discrepancy Resolution}: In instances where the AI findings diverged from the initial human assessment, the specific points of contention were re-evaluated. This step involved a detailed review of the relevant literature and framework documentation to resolve discrepancies, thereby refining the accuracy of our analysis.
	
	\item \textit{Final Validation}: After achieving consistency between human and AI assessments, the final validation was conducted via expert review to confirm the robustness of the findings. This multi-stage process ensured a rigorous evaluation, minimizing subjective bias and enhancing the reliability of the qualitative analysis.
\end{itemize}

\section{Results}
\label{sec:results}
This section presents the findings from our analysis (Table \ref{tab:AssessmentLLM}), showcasing automation potential across frameworks, their capabilities for overseeing LLM-related risks, and identified gaps in readiness for LLM adoption and integration.

\begin{table*}[t!]
	\centering
	\caption{Assessment of LLM opportunities, risks and EU AI Act compliance (\checkmark: pass; $\times$: fail)}
	\label{tab:AssessmentLLM}
	\resizebox{\textwidth}{!}{%
		\begin{tabular}{lcccccc}
			\hline
			& & & \textbf{NIST CSF 2.0} & \textbf{COBIT 2019} & \textbf{ISO 27001:2022} & \textbf{ISO 42001:2023} \\ \hline
			\multirow{3}{*}{\textbf{Process automation}} & \multirow{2}{*}{LLM} & opportunities & $\times$ & $\times$ & \checkmark & \checkmark  \\ \arrayrulecolor{gray!40} \cline{3-7} 
			&  & risks & $\times$ & $\times$ & $\times$ &  $\times$ \\ \cline{2-7} 
			& \multicolumn{2}{c}{EU AI Act readiness} & $\times$ & \checkmark & $\times$ & \checkmark  \\ \hline
			
			\multirow{3}{*}{\textbf{Real-time analysis}} & \multirow{2}{*}{LLM} & opportunities & \checkmark & \checkmark & \checkmark & \checkmark \\ \cline{3-7} 
			&  & risks & $\times$ & \checkmark & $\times$ & $\times$ \\ \cline{2-7} 
			& \multicolumn{2}{c}{EU AI Act readiness} & $\times$ & \checkmark & $\times$ & \checkmark  \\ \hline
			
			\multirow{3}{*}{\textbf{\begin{tabular}[c]{@{}l@{}}Data security and \\ protection\end{tabular}}} & \multirow{2}{*}{LLM} & opportunities & \checkmark & \checkmark & \checkmark & \checkmark\\ \cline{3-7} 
			&  & risks & $\times$ & \checkmark & $\times$ & \checkmark \\ \cline{2-7} 
			& \multicolumn{2}{c}{EU AI Act readiness} & \checkmark & \checkmark & \checkmark & \checkmark \\ \hline
			
			\multirow{3}{*}{\textbf{\begin{tabular}[c]{@{}l@{}}Continuous monitoring \\ and auditing\end{tabular}}} & \multirow{2}{*}{LLM} & opportunities & \checkmark & \checkmark & \checkmark & \checkmark \\ \cline{3-7} 
			&  & risks & $\times$ & $\times$ & $\times$ & \checkmark \\ \cline{2-7} 
			& \multicolumn{2}{c}{EU AI Act readiness} & \checkmark & \checkmark & \checkmark &  $\times$ \\ \hline
			
			\multirow{3}{*}{\textbf{Incident response}} & \multirow{2}{*}{LLM} & opportunities & \checkmark & \checkmark & \checkmark & \checkmark \\ \cline{3-7} 
			&  & risks & $\times$ & $\times$ & $\times$ & \checkmark \\ \cline{2-7} 
			& \multicolumn{2}{c}{EU AI Act readiness} & $\times$ & \checkmark & $\times$ & $\times$ \\ \hline
			
			\multirow{3}{*}{\textbf{\begin{tabular}[c]{@{}l@{}}Security awareness\\  and training\end{tabular}}} & \multirow{2}{*}{LLM} & opportunities & $\times$ & \checkmark & \checkmark &  \checkmark \\ \cline{3-7} 
			&  & risks & $\times$ & $\times$ & \checkmark &  \checkmark \\ \cline{2-7} 
			& \multicolumn{2}{c}{EU AI Act readiness} & $\times$ & $\times$ & $\times$ & $\times$ \\ \hline
			
			\multirow{3}{*}{\textbf{\begin{tabular}[c]{@{}l@{}}Policy and compliance \\ checks\end{tabular}}} & \multirow{2}{*}{LLM} & opportunities & $\times$ & \checkmark & \checkmark & \checkmark \\ \cline{3-7} 
			&  & risks & \checkmark & $\times$ & \checkmark & $\times$  \\ \cline{2-7} 
			& \multicolumn{2}{c}{EU AI Act readiness} & $\times$ & \checkmark & \checkmark & $\times$ \\ \arrayrulecolor{black} \hline
			
			\multirow{3}{*}{\textbf{TOTAL MARKS}} & \multirow{2}{*}{LLM} & opportunities & 5/7 & 6/7 & \cellcolor{yellow!30}7/7 & \cellcolor{yellow!30}7/7 \\ \arrayrulecolor{gray!40} \cline{3-7} 
			&  & risks & 1/7 & 2/7 & 2/7 & \cellcolor{yellow!30}4/7 \\ \cline{2-7} 
			& \multicolumn{2}{c}{EU AI Act readiness} & 2/7 & \cellcolor{yellow!30}6/7 & 3/7 & 4/7 \\ \arrayrulecolor{black} \hline
		\end{tabular}%
	}
\end{table*}

\subsection{LLM opportunities}
\label{subsec:automation_opportunities}
To systematically evaluate the potential of each framework for LLM automation and augmentation, we employed the mapping rubric to identify compatible processes and controls. Our assessment indicated all three frameworks accommodated some aspects of LLM capabilities, with CSF 2.0 offering the most comprehensive set of opportunities due to its breadth of technical and governance outcomes. However, high-level alignment did not preclude the need for additional LLM-specific provisions to ensure responsible and risk-aware integration.

\paragraph{NIST CSF 2.0 (rating 5/7)}
The draft NIST CSF 2.0 exhibits potential for integration with LLM technologies within the domains of \enquote{real-time analysis}, \enquote{data security and protection}, \enquote{continuous monitoring and auditing}, and \enquote{incident response}. These areas, though not explicitly addressing LLMs, provide a conducive framework for their application—real-time analysis (\textit{ID.RA}), data protection (\textit{PR.DS}), continuous monitoring (\textit{DE.CM}), and incident response (\textit{RS.CO}). However, it falls short in \enquote{process automation}, \enquote{security awareness and training}, and \enquote{policy and compliance checks}, lacking specific references or provisions for LLM utilization, such as \textit{Natural Language Processing} (NLP) for automation, adaptive security training modules, and automated policy drafting and compliance verification. These gaps suggest that while the framework may be adaptable to LLM integration, it currently does not offer explicit readiness in these critical areas.

\paragraph{COBIT 2019  (rating 6/7)}
COBIT 2019 presents notable alignment with LLM opportunities in the areas of real-time analysis (\textit{APO12}), data security and protection (\textit{APO01}, \textit{BAI09}), continuous monitoring and auditing (\textit{APO11}), incident response (\textit{APO13}, \textit{DSS04}, \textit{DSS05}), security awareness and training (\textit{BAI08}), and policy and compliance checks (\textit{EDM01}, \textit{EDM02}, \textit{BAI01}, \textit{BAI02}), though it has not specified the use of LLMs within these provisions. The framework, however, does not pass in the domain of \enquote{process automation}, as it lacks explicit guidance on integrating LLMs for NLP or automation. COBIT 2019 does not provide specific references to leveraging LLM capabilities for process automation, highlighting a gap that may need to be addressed to fully harness LLM opportunities in this aspect.

\paragraph{ISO 27001:2022  (rating 7/7)}
ISO27001:2022 demonstrates a readiness to leverage the capabilities of LLMs across various domains pertinent to cybersecurity and information security management. While the standard does not explicitly detail the integration of LLMs, its broad and thorough controls provide a robust foundation for the secure adoption and implementation of LLM technologies. Controls related to information security testing, system development, event logging, and incident management indicate an infrastructure that is conducive to incorporating LLMs into process automation, real-time analysis, and continuous monitoring. Additionally, the framework acknowledges the importance of security awareness and training, which can be enhanced through adaptive training modules potentially supported by LLMs. In \enquote{data security and protection}, ISO27001:2022 requires the secure management of information throughout its lifecycle, which is essential for LLMs handling sensitive data. Continuous monitoring and auditing controls imply a supportive environment for LLMs’ continuous learning processes, ensuring that their evolving capabilities remain within the realm of secure operations. Incident response controls align well with the use of LLMs for automated detection and reporting, facilitating timely and effective incident management. Overall, while ISO27001:2022 is not LLM-specific, its flexible and technology-agnostic approach allows it to remain relevant as new technologies emerge, indicating a strong potential for integration with LLM opportunities. The standard’s focus on information security management aligns with the inherent needs of LLMs, especially regarding the protection of data, which they process and generate. The framework provides extensive coverage that can be interpreted to support the secure introduction and utilization of LLMs within the constraints of its control sets.

\paragraph{ISO 42001:2023 (rating 7/7)}
ISO 42001:2023, while not explicitly designed for LLM integration, offers a framework that can support LLM capabilities in several key areas. For \enquote{process automation}, its focus on continuous improvement and risk management aligns with the needs for natural language understanding and generation, potentially offering a supportive environment. In \enquote{real-time analysis}, the standard's emphasis on continual improvement and adaptability may facilitate real-time data processing. For \enquote{data security and protection}, ISO 42001's comprehensive risk management approach could be conducive to integrating data analytics, image recognition, and generation. The standard's focus on \enquote{continuous monitoring and auditing} aligns well with the needs for continuous learning in LLMs. In \enquote{incident response}, the structured approach to risk assessment and treatment may support automated incident detection and reporting. The \enquote{security awareness training} aspect could benefit from the standard's emphasis on awareness and competence, potentially enabling the integration of adaptive training modules. Lastly, in \enquote{policy checks and compliance}, ISO 42001's structured approach to managing AI systems and risks may align with the requirements for automated policy drafting and checks, though explicit provisions for LLMs are not detailed. Overall, while ISO 42001:2023 does not specifically address LLMs, its principles and focus on AIMS suggest a supportive framework for their integration, warranting further exploration and application to determine its full compatibility.

\subsection{LLM risks}
\label{subsec:risk_oversight_capabilities}
To systematically evaluate the provisions of each framework for governing LLM-associated risks, we employed the mapping rubric to identify relevant controls and requirements.

\paragraph{NIST CSF 2.0 (rating 1/7)}
NIST CSF 2.0 offers firm guidance within \enquote{Policy and compliance checks} via the \textit{GV.PO} category, requiring robust policies for safeguarding data and technology. This, however, marks the extent of its explicit coverage of LLM-related risks, particularly in the specialized areas of content generation, real-time bias correction, data protection specific to LLM technology, continuous LLM data oversight, targeted LLM incident response, and LLM-focused security education.

The framework does not satisfactorily address the intricacies of LLM risk in \enquote{Process Automation}, failing to offer specific strategies for the perils arising from LLM-generated content. The \enquote{Real-time Analysis} component, while presenting relevant categories, falls short in providing concrete measures for the unique temporal and bias-related challenges associated with LLM outputs. Within \enquote{Data Security and Protection}, NIST CSF 2.0 establishes a broad defense but stops short of probing into the advanced threats LLMs pose, such as data breaches and deceptive visual content. Its lack of detailed guidance on the relevance and security of training data indicates a gap in ``Continuous Monitoring and Auditing'', essential for LLM-specific oversight. ``Incident Response'' protocols are not adequately calibrated for the specific issues of LLM misclassifications, potentially leading to ineffective response actions. Moreover, the section on ``Security Awareness and Training'' neglects to incorporate tailored instruction to counteract LLM-related risks, such as distorted realities and data profiling, despite an overall emphasis on the importance of role-oriented training initiatives.

\paragraph{COBIT 2019 (rating 2/7)}
COBIT 2019 meets LLM risk readiness criteria in the areas of ``Real-time analysis'' and \enquote{Data security and protection}, with relevant provisions being \textit{EDM03} and \textit{DSS05} respectively. These sections provide a foundation for addressing risks associated with real-time processing and data protection, which could extend to encompass the unique challenges posed by LLMs. Conversely, COBIT 2019 fails to adequately address LLM risk readiness in the domains of \enquote{Process automation}, \enquote{Continuous monitoring and auditing}, \enquote{Incident response}, \enquote{Security awareness and training}, and \enquote{Policy and compliance checks}. Its general IT governance and management objectives do not sufficiently cover the specific risks associated with automated content generation by LLMs, nor do they ensure that risk responses are specifically tailored for the challenges posed by LLMs, such as incident misclassification or the introduction of non-compliant rules. Furthermore, there is an absence of detailed guidance for the management and monitoring of LLM training data and the subtle needs of AI/ML-specific employee training content, which is crucial for maintaining an informed and prepared workforce in the face of evolving LLM risks.

\paragraph{ISO27001:2022 (rating 2/7)}
ISO27001:2022 has demonstrated a foundational readiness for \enquote{Security awareness and training} and \enquote{Policy and compliance checks}, under provisions \textit{A.7.2} for fostering security knowledge, and \textit{A.18.1} and \textit{A.18.2} for policy management, yet these lack explicit directives for LLM-specific risks. The framework has not adequately covered \enquote{Process automation}, with no tailored controls for automation risks inherent to LLMs in its Annex A, notably absent in sections addressing separation of environments (\textit{A.12.4.1}). For \enquote{Real-time analysis}, it has fallen short, missing explicit consideration for LLM-induced biases and latency. The \enquote{Data security and protection} provision, although robust in its scope (\textit{A.8.2.3} and \textit{A.14.1.2} among others), has failed to specifically safeguard against LLM-related risks like visual deception and forensic unreliability. Its \enquote{continuous monitoring and auditing} aspect has lacked directives on ensuring the ongoing relevance and integrity of the training data. In \enquote{Incident response}, its general incident management controls (\textit{A.16.1}) have not directly addressed the unique challenge of LLM misclassification risks. Consequently, while ISO27001:2022 has established a broad security and compliance framework, it still requires significant enhancement to directly confront the unique challenges posed by LLM technologies.

\paragraph{ISO 42001:2023 (rating 4/7)}
ISO 42001:2023 has demonstrated a mixed readiness for managing LLM-related risks. In our assessment, the standard passed in the domains of \enquote{Data security and protection}, \enquote{Continuous monitoring and auditing}, \enquote{Incident response}, and \enquote{Security awareness training}. It provided comprehensive guidelines for data management (including privacy and security implications), AI system logging, and promoting security knowledge (Controls B.7.2, B.7.3, and A.7.2). However, it failed in the areas of \enquote{Process automation}, \enquote{Real-time analysis}, and \enquote{Policy and compliance checks}. While it addressed general AI system risks, specific references to managing misleading content generation, real-time biases, or the introduction of non-compliant rules in AI systems (including LLMs) were lacking. The closest relevant provisions included data quality requirements and ensuring the responsible use of AI systems (Controls B.7.4, B.9.2), yet these did not fully cover the complex risks posed by LLMs, such as automated decision-making biases or the unique challenges of real-time AI system responses. This gap indicated a need for more detailed and explicit risk management strategies for LLM technologies within the standard.

\subsection{EU AI Act readiness}
\label{subsec:EU_AI_Act_readiness}
The EU AI Act draft introduces specific provisions for organizations implementing LLMs, which are distinct from those for regular AI systems due to the unique capabilities and risks associated with LLMs:

\begin{itemize}
	\item \textbf{Risk Management Systems (\textit{Article 9})}: LLMs can process and generate content at scale, increasing the risk of widespread misinformation or data manipulation. 
		\subitem Our recommendation: Implement systems to assess, document, and minimize such cybersecurity risks.
	
	\item \textbf{Mandatory Cybersecurity Testing (\textit{Article 15})}: The complexity and depth of LLMs may harbor hidden vulnerabilities.
		\subitem Our recommendation: Require extensive testing for vulnerabilities and data integrity before deploying LLMs.
	
	\item \textbf{Transparency Obligations (\textit{Article 13})}: LLMs' \enquote{black box} nature makes understanding their decision-making processes challenging.
		\subitem Our recommendation: Mandate documentation on high-risk LLMs' capabilities, limitations, and security measures for clarity.

	\item \textbf{Post-Market Monitoring (\textit{Article 61})}: The evolving nature of LLMs means new risks can emerge after deployment.
		\subitem Our recommendation: Require continuous monitoring for cybersecurity issues.

	\item \textbf{Record-Keeping (\textit{Article 11-12})}: The adaptive learning of LLMs necessitates detailed records of their design, risk assessments, and evaluations.
		\subitem Our recommendation: Make such records accessible for authority review to ensure ongoing compliance.

	\item \textbf{Reporting Obligations (\textit{Article 62})}: Given LLMs' potential impact, significant cyber incidents must be reported to authorities.
		\subitem Our recommendation: Ensure accountability and rapid response to threats posed by LLMs, and prompt reporting of serious incidents and malfunctioning to regulatory bodies.

	\item \textbf{Appointment of Cybersecurity Officers (\textit{Article 17})}: LLMs require specialized oversight due to their complex nature.
		\subitem Our recommendation: Appoint qualified cybersecurity officers to oversee LLM security compliance.

	\item \textbf{Fines for Non-Compliance (\textit{Article 71})}: Non-compliance with LLM-specific cybersecurity requirements can result in financial penalties.
		\subitem Our recommendation: Adhere to the heightened security needs of LLMs.
\end{itemize}

Given the three frameworks analyzed in relation to the EU AI Act provisions, none explicitly include AI-specific stipulations. However, they exhibit varying degrees of implicit alignment with the Act requirements, with some fulfilling numerous provisions without necessitating major amendments.

\paragraph{NIST CSF 2.0 (rating 2/7)}
The NIST CSF 2.0 has demonstrated alignment with the EU AI Act in areas of \enquote{data security and protection} (\textit{PR.DS}) and \enquote{continuous monitoring and auditing} (\textit{DE.CM}), emphasizing robust protection against data breaches, ensuring AI data quality and integrity, and fostering continuous monitoring with periodic AI assessment. The \enquote{process automation} provision of NIST CSF 2.0 is not EU AI Act ready, because it lacks specific requirements related to transparency, oversight, and compliance of automated processes, and may not fully address the AI-specific requirements. Its \enquote{real-time analysis} provision does not adequately cater to real-time safeguards and unbiased AI system responsiveness. Its \enquote{incident response} provision fails to specify rapid AI-driven incident recognition and strategies for AI misclassification risk mitigation. Its \enquote{security awareness and training} provision is deficient in terms of AI transparency in training, authenticity and accuracy of AI-focused training information, and profiling of employees. Lastly, the \enquote{policy and compliance checks} provision is not comprehensive in addressing automated information verification and direct guidelines for AI alignment with compliance.

\paragraph{COBIT 2019 (rating 6/7)}
COBIT 2019 has exhibited strong EU AI Act readiness across several provisions, with its governance and management objectives addressing requirements for process automation, real-time risk management, data security and protection, continuous monitoring and auditing, incident response, and policy and compliance checks, emphasizing transparency, oversight, real-time safeguards, robust data protection, continuous evaluation, agile incident responses, and compliance alignment. Its \enquote{security awareness and training} provision is not EU AI Act ready, because while it promotes comprehensive training and awareness related to ethics, transparency, and appropriate use of information, it may lack in-depth AI-specific considerations in alignment with the EU AI Act, potentially missing direct provisions on employee profiling in an AI context and specific guidelines on the authenticity and accuracy of AI-focused training information.

\paragraph{ISO 27001:2022 (rating 3/7)}
ISO 27001:2022 has demonstrated a degree of EU AI Act readiness in its provisions related to data security and protection, continuous monitoring and auditing, and policy and compliance checks, emphasizing robust guidelines against data breaches, continuous evaluation of information security controls, and comprehensive policy and compliance assessment. However, there are areas of concern: its \enquote{process automation} provision is not EU AI Act ready, as it lacks specific guidance around transparency, oversight, and compliance for automated processes in the AI context. Its \enquote{real-time analysis} provision does not fully address the requirements around unbiased AI system responsiveness and real-time safeguards. Its \enquote{incident response} provision, while robust in general incident management, does not target AI-driven recognition or misclassification risks, necessitating further guidelines to handle challenges posed by AI systems. Its \enquote{security awareness and training} provision lacks direct provisions for AI-specific issues such as AI transparency in training and employee profiling restrictions.

\paragraph{ISO 42001:2023 (rating 4/7)}
ISO 42001:2023 has demonstrated considerable readiness in several aspects of EU AI Act compliance, but with areas needing further enhancement. In \enquote{Process automation}, it aligns well with transparency and human oversight requirements (Article 52-53, 61, 63) through its focus on risk treatment and effectiveness verification (\textit{Clauses 6.1.3 and 6.1.4}). For \enquote{Real-time analysis}, ISO 42001:2023 partially meets the criteria of real-time safeguards and unbiased AI responsiveness (Article 5, 9, 62, 65) through its provisions for monitoring and measuring AIMS performance (\textit{Clause 9.1}). The framework effectively addresses \enquote{Data security and protection} with robust protection against data breaches and integrity of AI data (Article 15, 70, Annex IV-2, VII-4.3) by ensuring effective internal audits (\textit{Clauses 9.2 and 9.2.1}) and top management reviews (\textit{Clause 9.3}). However, gaps are observed in \enquote{Continuous monitoring and auditing} and \enquote{Incident response}, lacking direct provisions for periodic AI assessments and AI-driven incident recognition, despite general clauses on corrective actions and nonconformity management (\textit{Clause 10.1}). \enquote{Security awareness training} is partially covered, addressing AI transparency (\textit{Clause 7.4}), but lacking specifics on authenticity, accuracy, and profiling restrictions. In \enquote{Policy and compliance checks}, ISO 42001:2023 excels in automated information verification and AI compliance (Article 60, 61, 64, 8, 9, 16, 24-27), thanks to its comprehensive framework for AIMS implementation, maintenance, and continuous improvement, providing a structured approach to AI governance and compliance.

\subsection{Gap analysis}
\label{subsec:gap_analysis}
\begin{figure*}[t!]
	\centering
	\begin{tikzpicture}
		\begin{axis}[
			xbar stacked,
			bar width=20pt,
			width=0.85\textwidth,
			height=0.35\textwidth,
			xlabel={LLM readiness score per category (the higher, the better)},
			xmajorgrids=true,
			xtick={0,3,...,21}, 
			grid style=dashed,
			xmin=0,
			xmax=21,
			enlarge y limits={abs=0.7cm},
			symbolic y coords={ISO 42001:2023,ISO 27001:2022,COBIT 2019,NIST CSF 2.0},
			ytick=data,
			nodes near coords,
			nodes near coords align={horizontal},
			every node near coord/.append style={
				font=\small,
				/pgf/number format/precision=0,
				/pgf/number format/fixed zerofill=true,
				anchor=center
			},
			legend style={at={(0.45,-0.3)},anchor=center,legend columns=-1}
			]
			
			\addplot[fill=blue!30!white,postaction={pattern=north east lines, pattern color=black!50!white}] coordinates {(7,ISO 42001:2023) (7,ISO 27001:2022) (6,COBIT 2019) (5,NIST CSF 2.0) };
			\addplot[fill=red!30!white, postaction={pattern=vertical lines, pattern color=black!50!white}] coordinates {(4,ISO 42001:2023)(2,ISO 27001:2022) (2,COBIT 2019) (1,NIST CSF 2.0)};
			\addplot[fill=green!30!white, postaction={pattern=north west lines, pattern color=black!50!white}] coordinates {(4,ISO 42001:2023)(3,ISO 27001:2022) (6,COBIT 2019) (2,NIST CSF 2.0)};

			\legend{LLM opportunities (7)~~~~~~~, LLM risks (7)~~~~~~~, EU AI Act (7)}
		\end{axis}
	\end{tikzpicture}
	\caption{Comparison of cybersecurity GRC frameworks in LLM readiness}
	\label{fig:framework_comparison}
\end{figure*}
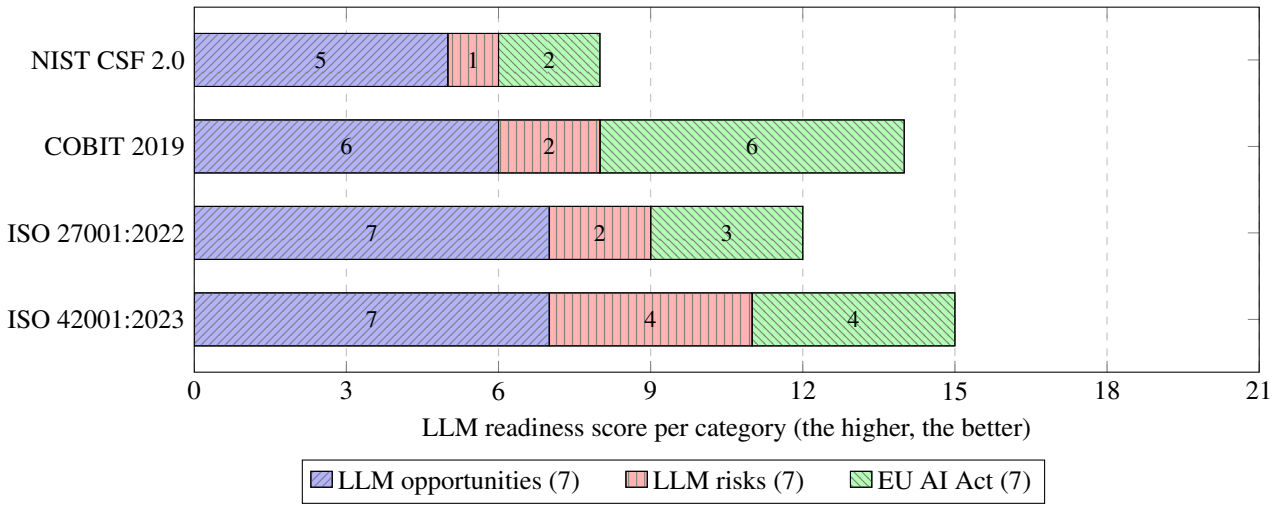

Our assessment has revealed key insights into the readiness of the frameworks, including the latest ISO 42001:2023, for LLM integration along two dimensions: automation potential and risk oversight. A comparative analysis highlights crucial gaps that need to be addressed as summarized in Table \ref{table:gap_analysis}. While the NIST CSF 2.0 offers extensive automation potential, it lacks explicit LLM risk oversight. COBIT 2019 facilitates high-level automation opportunities but requires more granular technical controls for LLM-specific risks. ISO 27001:2022 provides a solid foundation for human-centered LLM adoption, yet needs augmentation for full automation potential. ISO 42001:2023, although not specifically targeting LLMs, shows promise in several domains such as process automation and data security, but requires further refinement in areas like real-time analysis and policy compliance for LLM applications. Our findings emphasized the necessity for a multi-dimensional approach as cybersecurity frameworks evolve to support LLM integration. This involves addressing both automation opportunities and strengthening risk oversight specific to LLM technologies. All frameworks, including ISO 42001:2023, while showing automation readiness, need enhancement to implement LLMs securely. This could be through oversight processes for NIST CSF 2.0, technical validations for COBIT 2019, automation-focused provisions for ISO 27001:2022, and more explicit LLM-related guidelines in ISO 42001:2023. The findings reiterate the need for frameworks to adopt a multi-dimensional view (Fig. \ref{fig:framework_comparison}), considering automation potential, oversight, and EU AI Act readiness, to support the integration of LLMs into cyber risk management programs. Consequently, we recommend caution if organizations wish to adopt any existing cybersecurity GRC framework without developing a false sense of security in adopting LLM opportunity readiness, LLM risk readiness, and EU AI Act readiness.

\begin{table*}[t!]
	\centering
	\caption{Comparative gap analysis of cybersecurity frameworks in LLM readiness}
	\label{table:gap_analysis}
	\resizebox{\textwidth}{!}{%
		\begin{tabular}{|p{2.2cm}|p{5.5cm}|p{5.5cm}|p{5.5cm}|}
			\hline
			\textbf{Framework} & \textbf{LLM opportunities} & \textbf{LLM risks} & \textbf{EU AI Act} \\
			\hline
			NIST CSF 2.0 & Incomplete provisions for process automation, security training, and policy compliance specific to LLM; lacks references to LLM for natural language processing, adaptive security training, and automated policy compliance. & Insufficient measures for LLM-generated content risks, real-time bias, advanced data breach threats, LLM data oversight, and tailored LLM incident response. & Partially aligns with data security and monitoring but lacks readiness in process transparency, oversight, unbiased responsiveness, AI-driven incident response, and AI-specific training and compliance. \\
			\hline
			
			COBIT 2019 & Omits explicit guidance on LLM integration for process automation; though covering various domains, it misses out on natural language processing and automation specific to LLM. & Does not cover LLM automated content generation risks; lacks LLM-tailored risk responses and specific AI/ML training for workforce. & Exhibits substantial readiness; however, its security training lacks depth in AI-specific considerations, employee profiling in AI, and training authenticity as per EU AI Act. \\
			\hline
			ISO 27001:2022 & While broad, does not detail LLM integration; controls need interpretation to support LLM application in process automation, real-time analysis, and continuous monitoring. & Provides foundation for security training and policy compliance but lacks explicit LLM risk directives; inadequate for LLM automation, real-time analysis, and incident misclassification risks. & Addresses some EU AI Act requirements but is not ready in providing guidance for transparency in AI process automation, real-time unbiased AI responses, and AI-driven incident management specifics. \\
			\hline
			
			ISO 42001:2023 & Supports LLM capabilities in process automation, real-time analysis, data security and protection, and more, through its focus on AIMS; however, does not explicitly detail LLM integration. & Effective in certain domains like data security and incident response, but lacks explicit strategies for LLM-specific risks such as misleading content generation and real-time biases. & Shows considerable readiness in some aspects of the EU AI Act, aligning well with transparency and oversight requirements, but has gaps in continuous monitoring and AI-driven incident response specifics. \\
			\hline

		\end{tabular}
	}
	\label{tab:LLM_Frameworks}
\end{table*}

\subsection{Common weakness in addressing LLM hallucination}
\label{subsec:hallucination_weakness}
The oversight of \enquote{misleading content generation}, colloquially known as LLM \enquote{hallucination} (\cite{ji2023survey,mcintosh2023google,mcintosh2023culturally}), emerged as a shared deficiency across the NIST CSF 2.0, COBIT 2019, ISO 27001:2022, and ISO 42001:2023 frameworks. This includes culturally sensitive hallucinations, which are particularly challenging given their dependency on cultural context and norms. Understanding this oversight necessitates a deep dive into the complex nature of hallucination and its implications for cybersecurity. Existing literature defines LLM hallucination as text generated by LLMs that contains factual inconsistencies, contradictions, or content that diverges from human cultural norms and expectations, even when being coherent and seemingly realistic \cite{ji2023survey}. This definition, while capturing the essence of the problem, falls short in addressing the subjective nature of hallucinations, particularly those that are culturally sensitive as highlighted in \cite{mcintosh2023culturally,mcintosh2024inadequacies}. To address this, our definition of LLM hallucination has been expanded to include not only factually inconsistent outputs but also those outputs that might be culturally incoherent or diverge from mainstream human values and expectations. This broader perspective recognizes that identifying a hallucination is not merely a task of matching facts but also involves the application of cultural and value-based perspectives, emphasizing the importance of human-expert-centric assessments. From a cybersecurity perspective, the implications of LLM hallucinations, including culturally sensitive ones, are multi-dimensional:

\begin{enumerate}[label=\arabic*)]
	\item \textit{Misinformation and Disinformation}: LLM hallucination poses risks of propagating misinformation and disinformation, both general and culturally specific. In cybersecurity processes, relying on culturally inappropriate or factually incorrect information can lead to flawed decision-making.
	
	\item \textit{Integrity of Data}: Compromised data quality and reliability due to LLM hallucinations can lead to erroneous conclusions in cybersecurity decision-making.
	
	\item \textit{Diverse Stakeholder Impact}: Cybersecurity involves stakeholders from various backgrounds. Culturally incoherent hallucinations could lead to misinterpretations or be considered offensive, affecting collaboration and trust.
	
	\item \textit{Decision-making Complexity}: Complex risk assessments in cybersecurity are further complicated by hallucinated LLM outputs, which could lead to decision-making paralysis.
	
	\item \textit{Human-AI Dynamics}: The rise of LLM-driven tools in cybersecurity necessitates harmony between human decisions and LLM recommendations. Hallucinations, especially those that are culturally sensitive, could challenge human-AI collaboration.
\end{enumerate}

The absence of adequate provisions to address LLM hallucinations, including culturally sensitive ones, in the frameworks has highlighted a broader issue: while these standards are forward-looking, they might not yet be equipped to address the complex challenges posed by emerging technologies like LLMs. They require frequent and robust updates to integrate specific checks and controls to identify, manage, and mitigate the risks posed by LLMs, including the culturally sensitive aspects of hallucinations.

\section{Discussion}
\label{sec:discussion}
The integration of LLMs into the realm of cybersecurity frameworks presents a multi-dimensional challenge that intersects both technological capabilities and governance oversight. In this section, we present some of our insights for discussion.

\subsection{The necessity of multi-pronged framework evolution}
Our findings have underlined the need for cybersecurity frameworks to undergo a multifaceted evolution, incorporating insights from the recently introduced ISO 42001:2023. Just as previous research identified gaps in handling emerging technologies like cloud computing~\cite{darraj2019artificial} and general AI~\cite{kaur2023artificial}, our analysis extends these insights to include LLMs. Current frameworks, including the newly added ISO 42001:2023, show foundational readiness but require enhancements for fully harnessing LLM capabilities and effectively managing their inherent risks. This evolution is in line with the dual-track approach that advocates for balancing innovation and oversight in emerging technology adoption~\cite{asad2023human,ukil2023knowledge}. ISO 42001:2023, while designed for and comprehensive in AI system management, still demonstrated areas needing refinement, particularly in addressing LLM-specific risks and compliance checks. This echoed the necessity of framework evolution to encompass adaptive controls tailored to the unique opportunities and challenges posed by LLMs~\cite{ji2023survey,darraj2019artificial}. Similarly, NIST CSF 2.0 and ISO 27001:2022, despite their foundational strengths, require updates to capitalize fully on LLM integration and risk management. This includes NIST CSF 2.0 enhancing its risk oversight, particularly for mitigating misleading content, and ISO 27001:2022 expanding its guidance for leveraging automation in training and monitoring. COBIT 2019's need for additional technical provisions for LLM-specific risk management reiterates this trend.

Cybersecurity GRC frameworks must refine their guidelines on training, system development, automation, and policy compliance, now also considering the guidelines established by ISO 42001:2023, to fully unlock LLMs' potential in transforming cybersecurity operations~\cite{akande2023cybersecurity,abie2019cognitive,markopoulou2019new}. The integration of human-centered collaboration, involving a diverse range of stakeholders, into framework design and implementation is essential. This ensures that frameworks reflect the priorities of both technology leaders and ethical oversight experts, thereby enhancing real-world efficacy~\cite{paskauskas2022enisa,malaivongs2022cyber,manuel2022cybertomp}. Our findings, along with recent scholarship, advocate for a re-examination of risk paradigms in light of AI advancements, acknowledging that current models may not fully account for the dynamic nature of technologies like LLMs~\cite{katina2021complex,manuel2022cybertomp}. Thus, organizations must adopt a proactive stance in framework implementation, anticipating and aligning with the evolving capabilities and risks of LLMs to maintain cybersecurity effectiveness~\cite{karie2021review,kaur2023artificial,gupta2023chatgpt,ukil2023knowledge}.

\subsection{The significance of continuous evolution and version control}
Our analysis found that frameworks must undergo rapid yet robust evolution to address emerging technologies. However, version control is crucial to ensure organizational adoption keeps pace with framework revisions. The identified gaps in the latest versions of the NIST CSF, COBIT, and ISO frameworks concerning LLM oversight underlined concerns about the frameworks' agility in keeping up with AI advances, highlighted in studies on framework modernization challenges \cite{gourisetti2020cybersecurity,hitchcox2020limitations}. While the pace of technological change is a reasonable challenge, it necessitates urgent version updates coupled with effective transition planning. This is to minimize prolonged lapses in readiness, a critical point underlined by researchers studying the adaptability and responsiveness of cybersecurity frameworks to emerging threats \cite{gourisetti2020cybersecurity,hitchcox2020limitations}. Our findings complement this discourse by demonstrating that the limitations of existing frameworks extend beyond operational aspects; they are conceptual, often failing to incorporate anticipatory governance necessary for technologies like LLMs~\cite{katina2021complex,manuel2022cybertomp}. This concept echoes the necessity for organizations not only to update their GRC frameworks more frequently, but also to integrate forward-looking approaches that can keep pace with AI innovation~\cite{darraj2019artificial,kaur2023artificial}. Therefore, we advocate for strategic version control to ensure that updated frameworks permeate through organizational infrastructure in a timely manner.

Our findings have thus highlighted that continuous evolution of frameworks must be complemented by responsible version release and adoption within organizations. All three frameworks need enhanced evolution to address technological changes rapidly, paired with strategic organizational implementation of updated versions. This emphasizes the significance of agile development and timely adoption for cybersecurity frameworks to remain relevant against emerging technologies. Balancing evolution and adoption is key for frameworks to continue fulfilling their vital role as cyber risk navigation tools in a climate of unrelenting change \cite{angelini2017crumbs}.

\subsection{Strengthening provisions for LLM hallucination risks}
Our investigation has exposed a lack of human expert (not simply human) oversight in the management of LLM hallucination risks within the NIST CSF 2.0, COBIT 2019, ISO 27001, and ISO 42001 frameworks, a concern that may be prevalent across other cybersecurity frameworks. This absence of controls aligns with the discourse in prior works advocating for risk management strategies tailored to AI's unique threats~\cite{darraj2019artificial,karie2021review,kaur2023artificial}, which our focus on LLM hallucination risks specifically seeks to advance. The deceptive nature of LLM hallucinations, which can be both subtle and overt, exacerbates these risks, especially when paired with human complacency or insufficient human oversight \cite{ji2023survey}. Those frameworks have been found to lack LLM-specific controls, particularly against the propagation of misleading content, which poses risks such as misinformation spread, data integrity breaches, stakeholder misalignment, decision-making disruption, and compromised human-AI collaboration. Prior research has underlined the necessity for cybersecurity measures that specifically address the unique threats posed by AI, advocating for a shift in risk management strategies to further encompass LLM's distinct threat profile~\cite{darraj2019artificial,karie2021review,kaur2023artificial,argyridou2023cyber,bozkus2018cyber}. To bridge these gaps, organizations must adopt a strategic approach that recognizes LLM as an autonomous entity within the threat landscape, and tactically integrate LLM risk scenarios into their cybersecurity exercises to fine-tune their response to LLM-specific threats~\cite{katina2021complex,manuel2022cybertomp,syafrizal2020analysis}.

To enhance the human oversight of LLM hallucination risks, several measures are recommended for incorporation into the NIST, COBIT, and ISO frameworks. Instituting mandatory hallucination identification processes, such as confidence scoring and uncertainty quantification, can preemptively detect misleading LLM outputs~\cite{huang2023towards,schuster2022confident}. Implementing human validation checkpoints ensures critical human oversight in the review of LLM outputs, while mandated transparency around LLM training data and model functionality aids in discerning unreliable outputs~\cite{gupta2023chatgpt,mcintosh2023harnessing}. Continuous bias testing is also essential, uncovering and correcting distortions in LLM knowledge bases that may lead to hallucinations~\cite{mesko2023imperative}. By embedding these targeted provisions, the frameworks can significantly bolster organizational defenses against LLM hallucinations, offering a comprehensive model for security standards that aim to integrate LLMs and manage their complex vulnerabilities effectively.

\subsection{Limitations of the study}
This study has provided critical qualitative insights into the readiness of key cybersecurity frameworks to integrate advanced AI systems such as LLMs, yet it is important to recognize its limitations. The inherent subjectivity of qualitative content analysis introduces the risk of bias, which we sought to minimize through rigorous validation with LLMs with NLP and human GRC experts, although this does not allow for statistically generalizable conclusions and confines the findings to the specific context and datasets examined. The goal and depth of our investigation necessitated a focus on three principal frameworks, thus excluding other standards that might have yielded further comparative perspectives; however, this limitation was offset by the thorough examination possible within the selected frameworks. Additionally, the analysis was based solely on publicly available documents, which, while ensuring transparency and avoiding sampling bias, did not account for the subtleties of dynamic stakeholder interactions. Despite these constraints, the study has laid a substantial groundwork for ongoing research into the influence of LLMs on cybersecurity governance, risk, and compliance, highlighting the complex challenges and opportunities these advanced AI systems present. Furthermore, the study's \enquote{pass/fail} approach as a qualitative method, though common in compliance assessments to provide clear outcomes and remediation steps when necessary, introduces another layer of limitation by simplifying the subtle continuum of compliance into binary outcomes. The abstract and principle-based nature of cybersecurity GRC frameworks also contribute to the subjectivity of the analysis, akin to different legal interpretations of the same legislation. Nonetheless, our method of including both LLMs and human experts in the loop has aimed to minimize interpretive discrepancies.

\section{Future directions and implications}
\label{sec:future_directions}
Future work could utilize mixed methods, expanded scope, and cross-disciplinary perspectives to further enrich understanding of how leading cybersecurity frameworks can continue evolving to support the safe, ethical and effective adoption of rapidly emerging technologies like AI. Findings would inform development of agile, holistic and evidence-based standards and programs for cybersecurity governance, risk management and compliance. Here are some suggestions for future work based on the limitations and findings of this study:

\begin{itemize}
	\item Launch a survey targeting cybersecurity professionals to statistically quantify the readiness and identify gaps in LLM integration across sectors, aiming for data that can validate findings and guide framework updates.
	
	\item Include a broader range of cybersecurity frameworks like CIS Controls in the scope to achieve a more detailed comparative analysis and understand sector-specific requirements for AI systems.
	
	\item Undertake case studies to observe the practical application of frameworks in organizations, focusing on effectiveness and real-world challenges in managing AI risks, with an emphasis on the use of LLMs.
	
	\item Extend the investigation to other cutting-edge technologies, including IoT, for a holistic view of how current frameworks can adapt to the broader technological landscape.
	
	\item Evaluate the responsiveness of frameworks to the rapidly evolving landscape of AI-enhanced threats, emphasizing the need for swift integration of new protective measures.
	
	\item Conduct a thorough examination of how cybersecurity frameworks currently align with not just the EU AI Act, but also other emerging legislations and standards in AI ethics and security.
	
	\item Facilitate focus groups or utilize Delphi methods to dynamically extract expert insights, allowing for a richer, contextually subtle understanding of framework application in the era of generative AI.
	
	\item Probe the potential benefits of integrating cybersecurity frameworks with other disciplinary perspectives, such as ethics and psychology, to create a more robust approach to AI challenges.
	
\end{itemize}

\begin{figure*}[t!]
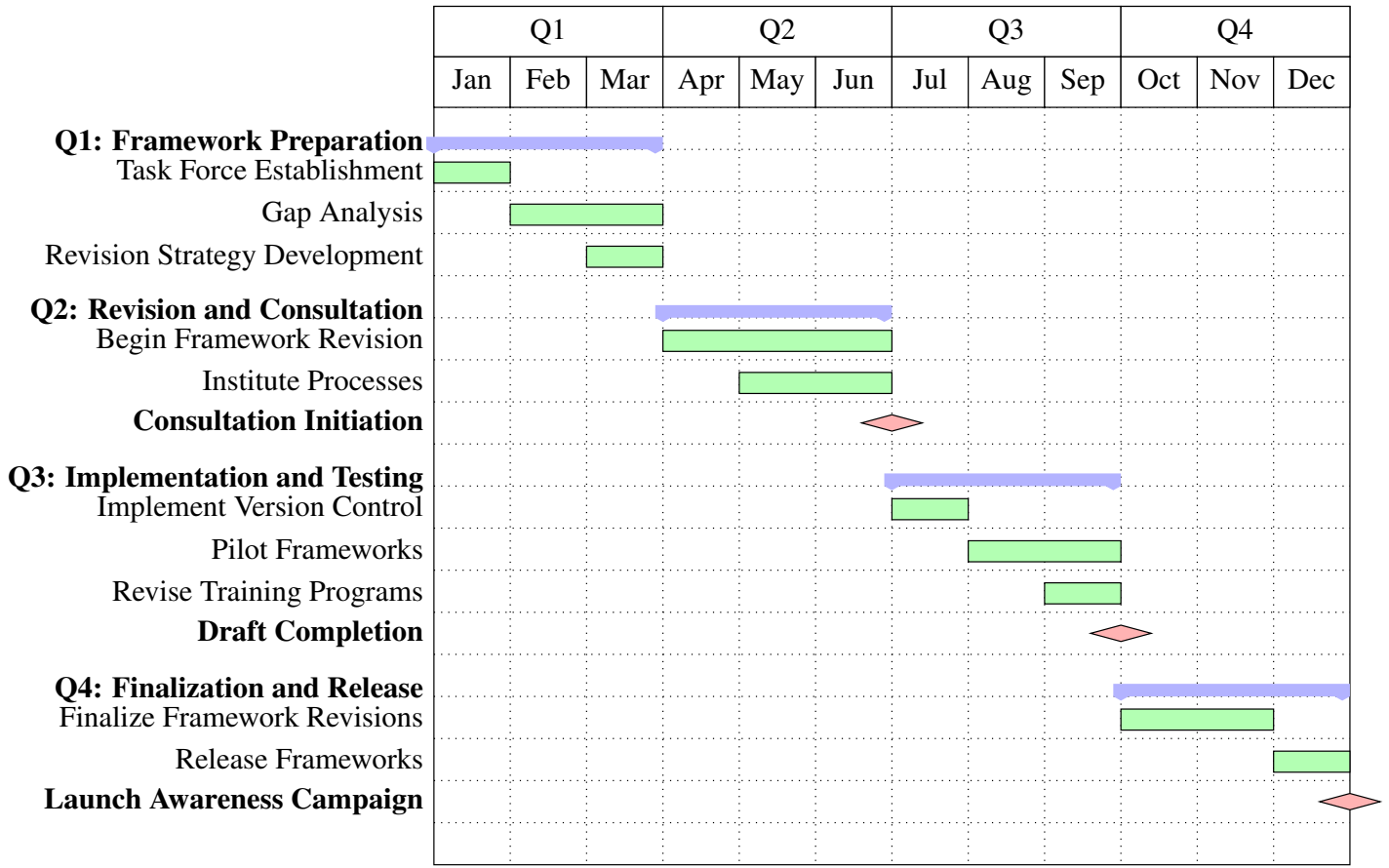

	\centering
	\resizebox{\textwidth}{!}{%
		\begin{ganttchart}[
			hgrid,
			vgrid,
			x unit=0.9cm,
			y unit title=0.6cm,
			y unit chart=0.5cm,
			title height=1,
			bar height=0.5,
			group right shift=0,
			group top shift=0.7,
			group height=.3,
			group peaks width={0.2},
			milestone label font=\bfseries\normalsize,
			bar label font=\normalsize,
			title label font=\normalsize,
			group/.append style={fill=blue!30},         
			bar/.append style={fill=green!30},          
			milestone/.append style={fill=red!30}
			]{1}{12}
			\gantttitle{Q1}{3} \gantttitle{Q2}{3} \gantttitle{Q3}{3} \gantttitle{Q4}{3}\\
			\gantttitle{Jan}{1} \gantttitle{Feb}{1} \gantttitle{Mar}{1} 
			\gantttitle{Apr}{1} \gantttitle{May}{1} \gantttitle{Jun}{1} 
			\gantttitle{Jul}{1} \gantttitle{Aug}{1} \gantttitle{Sep}{1} 
			\gantttitle{Oct}{1} \gantttitle{Nov}{1} \gantttitle{Dec}{1} 
			\\
			
			\ganttgroup{Q1: Framework Preparation}{1}{3} \\
			\ganttbar{Task Force Establishment}{1}{1} \\
			\ganttbar{Gap Analysis}{2}{3} \\
			\ganttbar{Revision Strategy Development}{3}{3} \\

			\ganttgroup{Q2: Revision and Consultation}{4}{6} \\
			\ganttbar{Begin Framework Revision}{4}{6} \\
			\ganttbar{Institute Processes}{5}{6} \\
			\ganttmilestone{Consultation Initiation}{6} \\

			\ganttgroup{Q3: Implementation and Testing}{7}{9} \\
			\ganttbar{Implement Version Control}{7}{7} \\
			\ganttbar{Pilot Frameworks}{8}{9} \\
			\ganttbar{Revise Training Programs}{9}{9} \\
			\ganttmilestone{Draft Completion}{9} \\

			\ganttgroup{Q4: Finalization and Release}{10}{12} \\
			\ganttbar{Finalize Framework Revisions}{10}{11} \\
			\ganttbar{Release Frameworks}{12}{12} \\
			\ganttmilestone{Launch Awareness Campaign}{12} \\

		\end{ganttchart}
	}
	\caption{1-Year plan Gantt chart to revise cybersecurity GRC frameworks}
	\label{fig:gantt}
\end{figure*}

Based on our insights, we also propose a year-long plan for responsible organizations in charge of those cybersecurity GRC frameworks (\textit{i.e.}, NIST CSF 2.0, COBIT 2019, ISO27001:2022, and ISO42001:2023), divided into four quarters, with the objective of updating the three cybersecurity GRC frameworks for the integration and regulation of LLMs. An illustrated Gantt Chart is provided in Fig. \ref{fig:gantt}. While acknowledging the complexity of revising cybersecurity GRC frameworks, and the possibility that such a task may extend beyond one year, the impending passage of the EU AI Act within the forthcoming year (2024) necessitates an expedited timeline. Consequently, we have designed this roadmap as a one-year project to address the criticality of the situation. Nevertheless, the project committee may exercise discretion to scale the pace of updates as required. 

\begin{itemize}
	\item {Quarter 1}
		\begin{itemize}
		\item Establish an interdisciplinary task force with expertise in cybersecurity, AI, and legal compliance.
		\item Conduct a comprehensive gap analysis to determine current framework deficiencies with respect to LLM integration and EU AI Act requirements.
		\item Develop a revision strategy for each framework, focusing on automation opportunities, risk governance, and regulatory compliance.
		\end{itemize}
	\item {Quarter 2}
		\begin{itemize}
			\item Begin framework revision with a focus on identifying and mitigating LLM hallucination risks.
			\item Institute processes for enhanced transparency, validation mechanisms, and bias testing specific to LLM usage.
			\item Initiate consultations with industry and academia to ensure the practicality and relevance of the proposed revisions.
		\end{itemize}
	
	\item {Quarter 3}
		\begin{itemize}
			\item Implement version control protocols to manage the transition to updated framework versions efficiently.
			\item Complete and pilot revised draft frameworks within selected organizations for real-world testing and feedback.
			\item Revise training programs and certification requirements to include LLM-specific content.
		\end{itemize}
	
	\item {Quarter 4}
		\begin{itemize}
			\item Finalize framework revisions, incorporating feedback from the pilot phase and ensuring alignment with the EU AI Act.
			\item Release the updated frameworks with comprehensive transition guidelines for organizations.
			\item Launch a global awareness campaign to inform stakeholders of the new revisions and encourage widespread adoption.
		\end{itemize}
\end{itemize}

\section{Conclusion}
\label{sec:conclusion}
This study has conducted a comprehensive evaluation of four leading cybersecurity frameworks - NIST CSF 2.0, COBIT 2019, ISO 27001:2022, and the recently introduced ISO 42001:2023 - in the context of their readiness for integrating and governing the rapidly evolving domain of Large Language Models (LLMs). Employing a detailed qualitative approach that includes content analysis, AI validation, and expert reviews, we unearthed critical insights that are of significant importance for both academic research and industry application in the realm of cybersecurity and generative AI. Our analysis has revealed a varying degree of alignment of these frameworks with the capabilities and risks associated with LLMs, indicating both promising potentials for integration and significant gaps in risk management. ISO 27001:2022 demonstrated strengths in human-centric validation for LLM outputs, reflective of its comprehensive approach to information security management. However, it became evident that all frameworks, including the NIST CSF 2.0 and COBIT 2019, necessitate further refinement to fully embrace the opportunities and address the cybersecurity risks introduced by LLMs from both technical and governance perspectives. Interestingly, ISO 42001:2023, despite being the most recent framework specifically designed for AI management, was found to have certain limitations in fully addressing the unique challenges and opportunities presented by LLM commercialization. While ISO 42001:2023 emerged as a frontrunner in our comparative analysis, its principles and guidelines, primarily tailored for general AI management, did not fully encapsulate the specific subtleties and requirements of LLM technologies. This gap highlighted the necessity for even the most contemporary frameworks to undergo continuous evolution, ensuring that they are not only in tune with general AI advancements but also adequately responsive to the distinct characteristics of LLMs. The prevalent oversight of LLM-related risks, particularly the phenomenon of 'hallucination' or misleading content generation, across all examined frameworks, underlines the complex and pressing challenges posed by advanced LLM systems. This study emphasizes the urgent need for the modernization of mainstream cybersecurity standards to effectively govern these emerging threats. A critical takeaway is the requirement for frameworks to incorporate LLM-tailored controls that emphasize transparency, human validation, bias testing, and continuous monitoring. As the landscape of cybersecurity and AI continues to evolve, our findings call for a proactive and dynamic approach in the development and adaptation of cybersecurity frameworks, ensuring their relevance and efficacy in the age of advanced LLMs and beyond.

Stepping back, this study has made key contributions to cybersecurity knowledge by offering one of the first rigorous academic examinations of leading cybersecurity frameworks’ readiness for LLM integration, revealing crucial gaps and pathways for evolution. For the academic audience, our approach has provided a methodological model for assessing technological impacts on cybersecurity practices. For industry practitioners, it has guided them towards prioritizing framework updates that incorporate provisions for transparency, human oversight, and bias testing of LLM outputs. Collectively, our insights urge the field to take a multi-dimensional, agile and collaborative approach in shaping cybersecurity governance for the LLM era. Rather than a concluding perspective, this work seeks to spark further exploration at the nexus of generative AI and cybersecurity risk management. Through ongoing synthesis of cross-disciplinary insights, academia and industry can jointly cultivate forward-looking standards and programs that effectively harness LLM’s opportunities while diligently governing its risks and compliance.

\section*{Abbreviations}
\begin{table}[H]
	\resizebox{\columnwidth}{!}{
		\begin{tabular}{ll}
			AI & Artificial Intelligence \\
			AIMS & Artificial Intelligence Management System \\
			COBIT & Control Objectives for Information and Related Technologies \\
			CSF & Cybersecurity Framework \\
			EU & European Union \\
			GDPR & General Data Protection Regulation \\
			GPT & Generative Pre-trained Transformers \\
			GRC & Governance, Risk and Compliance \\
			ISO & International Organization for Standardization \\
			LLM & Large Language Model \\
			ML & Machine Learning \\
			NIST & National Institute of Standards and Technology \\
			NLP & Natural Language Processing 
		\end{tabular}%
	}
\end{table}

\bibliographystyle{elsarticle-num} 
\bibliography{LLM_Cyber_Framework}

\end{document}